\documentclass[]{aa}

\def\apj{ApJ}

\def\mn{MNRAS}

\def\ueber#1#2{{\setbox0=\hbox{$#1$}%
  \setbox1=\hbox to\wd0{\hss$ #2$\hss}%
  \offinterlineskip
  \vbox{\box1\box0}}{}}

\def\lapprox{\,\lower 1mm \hbox{\ueber{\sim}{<}}\,}
\def\gapprox{\,\lower 1mm \hbox{\ueber{\sim}{>}}\,}

\makeatletter

\let\@internalcite\cite
\def\cite{\@ifstar{\citeyear}{\citefull}}
\def\citefull{\def\astroncite##1##2{##1 ##2}\@internalcite}
\def\citeyear{\def\astroncite##1##2{##2}\@internalcite}

\def\citeau{\def\astroncite##1##2{##1}\@internalcite}
\def\citen{\def\astroncite##1##2{##1 (##2)}\@internalcite}
\def\possesivcite{\def\astroncite##1##2{##1's (##2)}\@internalcite}

\def\@citex[#1]#2{\if@filesw\immediate\write\@auxout{\string\citation{#2}}\fi
  \def\@citea{}\@cite{\@for\@citeb:=#2\do
    {\@citea\def\@citea{; }\@ifundefined
       {b@\@citeb}{{\bf ?}\@warning
       {Citation `\@citeb' on page \thepage \space undefined}}%
{\csname b@\@citeb\endcsname}}}{#1}}
\def\@cite#1#2{#1\if@tempswa , #2\fi}
\def\@biblabel#1{}
\makeatother

\begin{document}

\title{Carbon burning in intermediate mass primordial stars}

\author{Pilar Gil--Pons\inst{1},
        Takuma Suda\inst{2,3},
        Masayuki Y. Fujimoto\inst{2} \and
        Enrique Garc\'{\i}a--Berro\inst{1,4}}

\titlerunning{Carbon burning in intermediate mass primordial stars}
\authorrunning{Gil--Pons et al.}

\institute{$^1$Departament    de   F\'\i    sica    Aplicada,   Escola
               Polit\'ecnica  Superior  de Castelldefels,  Universitat
               Polit\`ecnica de Catalunya, Avda. del Canal Ol\'\i mpic
               s/n,   08860  Castelldefels,  Spain,   (e-mail:  pilar,
               garcia@fa.upc.es)\\
           $^2$Department  of  Physics, Hokkaido  University, Sapporo, 
               Hokkaido 060--0810, Japan\\
           $^3$Meme Media Laboratory,   Hokkaido  University, Sapporo, 
               Hokkaido 060--0813, Japan\\
           $^4$Institute for Space Studies of Catalonia, c/Gran Capit\`a
               2--4, Edif.  Nexus 104, 08034 Barcelona, Spain}

\date{}

\abstract{The evolution of a zero  metallicity $9 \, M_{\sun}$ star is
computed, analyzed and compared with  that of a solar metallicity star
of identical ZAMS mass.  Our computations range from the main sequence
until  the formation of  a massive  oxygen-neon white  dwarf.  Special
attention has  been payed to  carbon burning in conditions  of partial
degeneracy as  well as to  the subsequent thermally  pulsing Super-AGB
phase.  The  latter develops in  a fashion very  similar to that  of a
solar  metallicity $9  \,  M_{\sun}$  star, as  a  consequence of  the
significant enrichment  in metals of the stellar  envelope that ensues
due  to the so--called  third dredge--up  episode.  The  abundances in
mass of  the main isotopes  in the final  ONe core resulting  from the
evolution  are $X(^{16}$O)$\approx  0.59$,  $X(^{20}$Ne)$\approx 0.28$
and $X(^{24}$Mg)$\approx 0.05$.  This core is surrounded by a $0.05 \,
M_{\sun}$ buffer mainly  composed of carbon and oxygen,  and on top of
it a He envelope of mass $\sim 10^{-4}\, M_{\sun}$.
\keywords{stars: evolution  --- stars:  abundances --- stars:  AGB ---
stars: interiors --- stars: white dwarfs}}

\maketitle

%_____________________________________________________________________

\section{Introduction}

The primordial  universe was mainly  composed of hydrogen  ($X \approx
0.77$), helium ($Y \approx 0.23$),  and some $^2$H, $^3$He and $^7$Li,
in a very small amount ---  the total abundance did not represent more
than  $10^{-10}$ of  the  total mass  of  the universe.   This set  of
abundances,  derived  from  the  {\sl standard  homogeneous  Big  Bang
nucleosynthesis}    (Alpher   \&    Herman,    1950;   Olive,   1999),
characterizes  the matter  out of  which the  first structures  of the
Universe  formed.  According  to the  accepted  hierarchical scenario,
these  large  structures  had  a  mass between  $10^5$  and  $10^8  \,
M_{\sun}$,  and very  soon after  their formation  became the  nest in
which primordial stars formed.

To this regard it is important  to mention at this point that there is
no  general agreement  on the  initial  mass function  (IMF) of  these
primordial stars.   Instead, this particular subject is  still today a
matter of strong debate.   Moreover, the original composition of these
stars was definitely determined  by the fragmentation and condensation
processes that  led to their  formation and, therefore, it  is closely
related  to the  adopted IMF.   The first  attempts on  this direction
pointed to  the dominance of very  massive stars.  The  reason was the
original lack of metallic molecules that allowed the cooling necessary
to form low  and intermediate mass objects.  Therefore,  a big deal of
the earliest theoretical work on  $Z=0$ stars was devoted to the study
of very massive objects ---  see, for instance, Ezer \& Cameron (1971)
and Carr et  al.  (1983).  However, as it was  claimed later, the mere
presence of  molecular hydrogen could allow the  cooling necessary for
low and intermediate mass objects  to be formed (Carlberg, 1981; Palla
et al., 1983;  Yoshii \& Saio, 1986).  This  result motivated the work
of D'Antona  (1982), Guenter  \& Demarque (1983)  and Fujimoto  et al.
(1984) on   intermediate   mass    primordial   stars.    The   latest
two-dimensional  simulations (Nakamura \&  Umemura, 2001),  have found
that a bimodal  IMF peaked at $\approx\, 2 \, M_{\sun}$  and at a mass
between 10 and $100  \, M_{\sun}$ could be appropriate.  Consequently,
the most recent publications keep studying both massive (Heger et al.,
2001;  Heger  \&  Woosley,  2000) and  low-intermediate  mass  objects
(Marigo et al.,  2001; Chieffi et al., 2001).   For instance, Heger et
al.  (2000) considered the evolution  of $Z=0$ stars of masses between
15 and $250\, M_{\sun}$, from their main sequence until core collapse.
They concluded that  massive objects in general were  important in the
production  of primordial  $^{14}$N and  that those  stars  within the
highest mass range  were the place were primordial Fe  and Ni could be
synthetized  --- see  also Goriely  \& Siess  (2002) and  Abia  et al.
(2001).  Using  the results of Limongi  et al.  (2000)  and Chieffi et
al. (2001) for Population III  stars, Abia et al.  (2001) analyzed the
impact  of  primordial stars  in  polluting  the intergalactic  medium
(IGM), and  showed the  relevance of  an IMF peaking  in the  range of
masses between  4 and $8  \, M_{\sun}$ in  order to explain  the large
[C/Fe] and [N/Fe] ratios observed in many extremely metal--poor stars.

In a pioneering  work on low and intermediate--mass  stars of very low
metallicity,   Fujimoto  et   al.    (1984),  studied   by  means   of
semianalytical  techniques  the  generation of  helium--shell  flashes
during the  AGB stage,  as a function  of mass and  metallicity.  They
found that  stars whose  ZAMS mass was  lower than $4\,  M_{\sun}$ did
experience  helium shell  flashes and  were important  contributors of
$^{12}$C and  neutron--rich isotopes of light  elements, whereas stars
with larger masses  were more likely to become  supernovae and, hence,
to contribute substantially to the production of Fe and Ni.  Marigo et
al.  (2001)  and Marigo  (2002) also studied  $Z=0$ stars in  the mass
range $0.7 \la M/M_{\sun} \la  100$, and followed their evolution from
the main  sequence until  the AGB  phase for objects  in the  low- and
intermediate-mass range.  In particular, they considered the pollution
of stellar  atmospheres and, ultimatelly, of  the interstellar medium,
as a  consequence of the  dredge--up processes ensuing during  the AGB
phase.   For the  high-mass range,  namely stars  with  masses $M_{\rm
ZAMS} \ga 9 \, M_{\sun}$, Marigo et al.  (2002) followed the evolution
of their model stars until  the carbon burning phase.  It is, however,
important to stress  at this point that in  the high-mass range carbon
is  burnt  in  non--degenerate  conditions.  Chieffi  et  al.   (2001)
focused on $Z=0$, $4 \la M/M_{\sun} \la 8$ stars and also followed the
thermally  pulsing phase  and  the dredge--up  during  the AGB  phase.
These  authors concluded  that  stars  in this  mass  range were  main
contributors of nitrogen and oxygen to the IGM.  Finally, Siess et al.
(2002) also followed  the evolution and  nucleosynthesis of primordial
stars  of masses  ranging between  0.8  and $20  \, M_{\sun}$.   Their
results, in agreement with those of Chieffi et al.  (2001), pointed to
a standard AGB behaviour for stars with masses larger than about $5 \,
M_{\sun}$.  These  authors have also  discussed the effects  of taking
into account the effects of a moderate amount of overshooting.

Finally,  it  is important  to  point out  that  one  might argue  the
scarcity of observational support  against the study of $Z=0$ objects.
In fact,  the detection of primordial composition  objects should take
us back to [Fe/H]$\approx -8.3$ epochs, whereas, up to now, the oldest
stars  detected are  a red  giant at  [Fe/H]$\approx -4.6$  (Bessel \&
Norris, 1984),  and a low mass star,  HE0107-5240, with [Fe/H]$\approx
-5.3$ (Christlieb  et al., 2002).  Nevertheless,  as correctly pointed
out by Chieffi et al.  (2001), the lack of observations should not let
us discard the existence of $Z=0$ intermediate mass stars.  Therefore,
in this paper  we study the evolution of  intermediate mass primordial
stars.

In order to  keep consistency with our previous  results, we have used
for the calculations reported in this paper the same evolutionary code
as that described  in Ritossa, Garc\'\i a--Berro \&  Iben (1995).  The
interested reader  can find there  a thorough description of  the most
relevant  physical  inputs   (neutrino  energy  loss  rates,  reaction
rates\ldots).  In  particular, the treatment  of convective boundaries
turns out to be important  for the calculations reported here. We have
not  adopted  any  overshoot  and  convective  regions  were  computed
according to the standard procedure  given by the mixing length theory
in  order to  adequately  compare with  the  calculations reported  in
Ritossa et al.  (1995). The  only major difference between the physics
used  there and those  used in  the present  calculations is  that, in
order to properly compute the evolution of $Z=0$ model star sequences,
we  have   implemented  the   corresponding  opacities  at   very  low
metallicities of Iglesias \& Rogers (1993).

The plan  of this paper is the  following: in section 2,  we study the
evolution of a  zero metallicity $9 \, M_{\sun}$  model star until the
beginning of carbon  burning, and we compare it  with the evolution of
the corresponding solar metallicity model star and with the results of
previous calculations.   In section 3,  the carbon burning  phase, the
associated thermonuclear flashes,  the dredge--up of nuclear processed
elements, and the associated changes in the surface composition of the
star  are  studied.   Section  4  is  devoted  to  the  study  of  the
thermally-pulsing  Super-AGB  (TP-SAGB)  phase.   In  section  5,  the
evolution  through the  post-AGB track  and  the final  phases of  the
evolution to become  a white dwarf are described.   Finally, section 6
is  devoted  to  summarize  our   main  results  and  to  outline  the
conclusions derived from them.

%____________________________________________________________________

\section{Hydrogen and helium burning}

\begin{table*}
\centering
\caption{Characteristics of  models at various points in the H--R  
diagram of Fig. 2}
\begin{tabular}{ccccccccccc} 
\hline
\hline
Model &  $t\, (10^{14}{\rm s})$ & $\log  L$ & $\log T_{\rm  eff}$ 
      &  $\log  R$  &  $\log\rho_{\rm c}$   &  $\log T_{\rm c}$ 
      & $M_{\rm He}$ & $X_{\rm c}$(He) & $M_{\rm C}$ & $X_{\rm c}$(C) \\ 
\hline 
A & 0.0000 & 3.63 & 4.55 & 0.25 & 4.72 & 7.71 & 0.00 & 0.23 & 0.00 &  0.00 \\
B & 2.3545 & 3.74 & 4.61 & 0.17 & 5.16 & 7.87 & 1.90 & 0.41 & 0.00 & $1\times 10^{-10}$ \\ 
C & 6.8994 & 3.95 & 4.53 & 0.44 & 5.24 & 7.93 & 1.90 & 1.00 & 0.00 & $3\times 10^{-9}$ \\
D & 7.0477 & 4.03 & 4.50 & 0.54 & 6.63 & 8.07 & 1.90 & 0.99 & 0.00 & $5\times 10^{-5}$  \\ 
E & 7.5843 & 4.10 & 4.53 & 0.55 & 6.34 & 8.22 & 2.03 & 0.48 & 0.52 & 0.45 \\ 
F & 8.0000 & 4.15 & 4.47 & 0.65 & 7.05 & 8.40 & 2.10 & 0.00 & 0.98 & 0.36 \\ 
G & 8.0520 & 4.08 & 4.14 & 1.29 & 8.80 & 8.64 & 2.15 & 0.00 & 1.05 & 0.36 \\ 
\hline
\hline
\end{tabular}
\end{table*}

We consider  a zero metallicity $9\,  M_{\sun}$ star.  As  we will see
below, the  characteristics of the evolution during  the main hydrogen
and helium  burning phases show important differences  with respect to
the  evolution   of  analogous   cases  of  higher   metallicity.   In
particular, as  already noted by  other authors that have  studied the
main sequence  phase of intermediate mass primordial  stars --- namely
Chieffi \& Tornamb\'e (1984), Tornamb\'e \& Chieffi (1986) and Cassisi
\& Castellani (1993) --- the first ascent to the red giant branch does
not take place before central  helium ignition but, instead, it occurs
when helium burning is set in a shell.

The main characteristic of $Z=0$ stars, that is, the absence of metals
and,  in particular, of  carbon, nitrogen  and oxygen,  determines the
evolution from its very early stages.  This composition does not allow
the CNO--cycle to ensue at  the beginning of the core hydrogen burning
phase,  as  it is  the  case for  intermediate  mass  stars of  higher
metallicity.  Instead, the pp--chains  must provide the nuclear energy
for the  star to  survive approximately during  the first  $7.4 \times
10^6$~yr.  However, the pp--chains are far from being the main nuclear
energy suppliers during  the bulk of core hydrogen  burning phase and,
as soon as  the temperature threshold for helium  burning reactions is
reached, the  production of $^{12}$C, $^{14}$N, and  $^{16}$O in small
amounts  (at abundance  thresholds of  about $10^{-10}$  in  mass), is
enough for the CNO--cycle to  take over as the primary energy supplier
during this  phase. It derives  from our calculations that  the entire
core hydrogen burning  phase lasts for about $2.1  \times 10^7$~yr, in
good agreement with  the result of Marigo et  al. (2001), who obtained
$2.2 \times 10^7$~yr.

\begin{figure}[t]
\vspace{7.9cm}
\hspace{-2.7cm}
\includegraphics{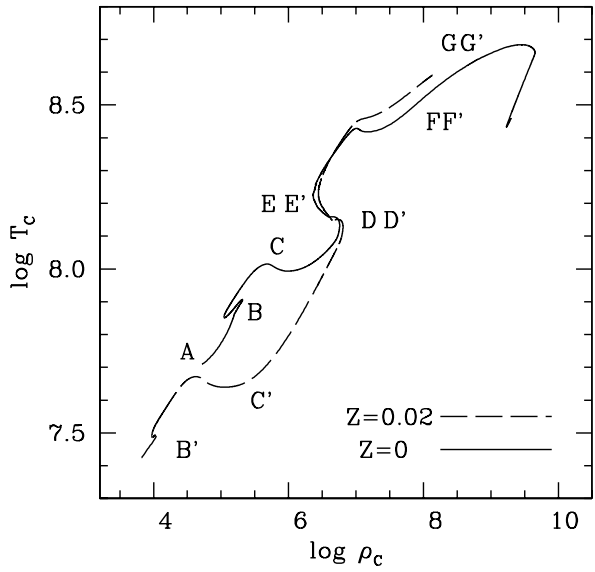}
\caption{Evolution in the $\log \rho_{\rm c} -\log T_{\rm c}$ plane of
        the zero metallicity $9\, M_\odot$  star (solid line) and of a
        $Z=0.02$ star of the same  mass (dashed line), up to the point
        where carbon is ignited.}
\end{figure}

\begin{figure}[t]
\vspace{7.9cm}
\hspace{-2.7cm}
\includegraphics{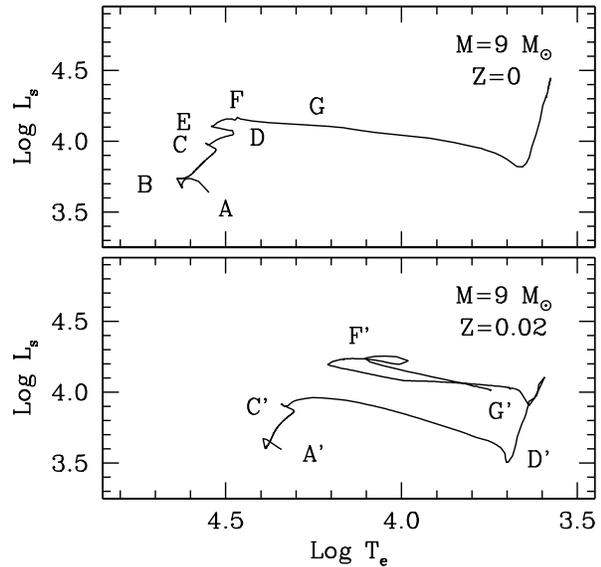}
\caption{Evolutionary  tracks in  the Hertzsprung--Russell  diagram of
        the $9\, M_{\sun}$ zero metallicity  star (top panel) and of a
        star  of $9\,  M_{\sun}$  with $Z=0.02$  (bottom panel).   The
        labels in the $Z=0$ model correspond to the models of Table 1.
        The  labels in  the  $Z=0.02$ model  correspond to  equivalent
        evolutionary stages.}
\end{figure}

The most important parameters  relevant for a correct understanding of
the  structure  and composition  of  our  models  during the  hydrogen
burning phase are  shown in Table 1.  For the  sake of completeness we
also  show  in  Table 1  the  same  parameters  for  the rest  of  the
evolutionary phases  before carbon is ignited.  The  positions of each
labelled model are  also shown in Fig.  1 and  Fig.  2, which display,
respectively, the  evolution of the central temperature  as a function
of the central  density for both our zero  metallicity $9 \, M_{\sun}$
star  and  that  of  a  $Z=0.02$  star  of  the  same  mass,  and  the
corresponding evolution  in the Hertzsprung--Russell  diagram for each
of  these  stars.    Models  labelled  as  A,  B   and  C  correspond,
respectively, to the beginning of the core hydrogen burning (hereafter
CHB),  to  the  time  at  which  the  energy  production  through  the
CNO--cycle starts to take over that of the pp--chains and, finally, to
the  end  of  CHB.   Models   labelled  as  D,  E  and  F  correspond,
respectively, to the  beginning, to an intermediate stage,  and to the
end  of central  helium burning  (hereafter CHeB).   Finally,  model G
corresponds to the time at which carbon is ultimately ignited.

\begin{figure}
\vspace{7.9cm}
\hspace{-2.7cm}
\includegraphics{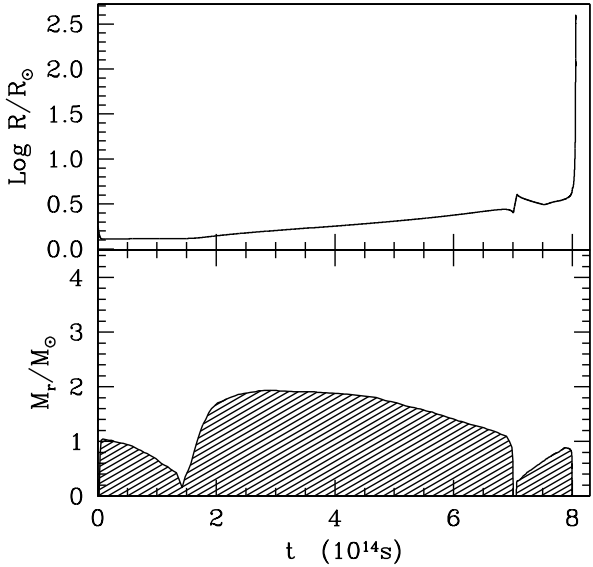}
\caption{Evolution of the convective regions of our model star (bottom
	panel) and of the radius of the star (top panel) as a function
	of time during the CHB and CHeB phases.}
\end{figure}

As it can be seen in Fig.  1, during the phase in which the pp--chains
are  the dominant  reactions  --- from  A  to B  --- core  contraction
proceeds accompanied by a  smooth increase in the central temperature.
However, once the  CNO--cycle is activated (at B),  the larger release
of energy  that consequently ensues temporarily  stops the contraction
and  lets  the  core  expand   before  the  usual  behaviour  of  core
contraction and heating during the  bulk of the CHB phase is attained.
For the  sake of comparison, we  have also represented in  Fig.  1 the
evolution  of the  central temperature  as a  function of  the central
density  for  the  case  $Z=0.02$.   As  it  can  be  seen,  important
differences arise.  The  loci in the $\log \rho-\log  T$ diagram where
core hydrogen burning begins  and fully develops are different because
hydrogen burning  through the pp--chains  requires higher temperatures
than  through  the CNO--cycle  to  achieve  the  same rate  of  energy
production.  This  fact also explains the positions  of the associated
loops in both  curves.  Both tracks are practically  coincident at the
point where the  3$\alpha$ reaction starts (D), as  central He burning
is not sensitive to  metallicity effects.  Nevertheless the subsequent
evolution takes the $Z=0$ star back to the region of higher densities.

The  evolution  in  the   Hertzsprung--Russell  diagram  of  the  zero
metallicity star also presents substantial differences with respect to
that  of  solar metallicity  (see  Fig.  2).   The  part  of the  core
hydrogen burning phase that proceeds through the pp--chains appears to
occur at slightly higher  effective temperatures and luminosities than
those of  the $Z=0.02$  case.  The bulk  of the core  hydrogen burning
phase,  occuring through  the CNO--cycle,  leads the  star to  a track
whose  shape is  very similar  to that  found for  the  $Z=0.02$ case.
Nonetheless, once this process is  completed, the star does not evolve
yet   to   the   low   $T_{\rm   eff}$--low   $L$   regions   of   the
Hertzsprung--Russell  diagram to  reach regions  close to  the Hayashi
track.   Instead, because  helium  burning occurs  very shortly  after
hydrogen  is exhausted  in the  core,  the star  keeps its  luminosity
almost unchanged,  and just a small decrease  in effective temperature
is observed.  The  ascent along the red giant branch  that can be seen
in Fig.  2 corresponds to  the helium--shell burning phase when, as it
will  be  shown  in  the  next section,  carbon  burning  has  already
developed in the  core. The former results are  in agreement with, for
instance, those of Marigo et al.  (2001) or Chieffi et al.  (2001).

\begin{figure}
\vspace{7.9cm}
\hspace{-2.7cm}
\includegraphics{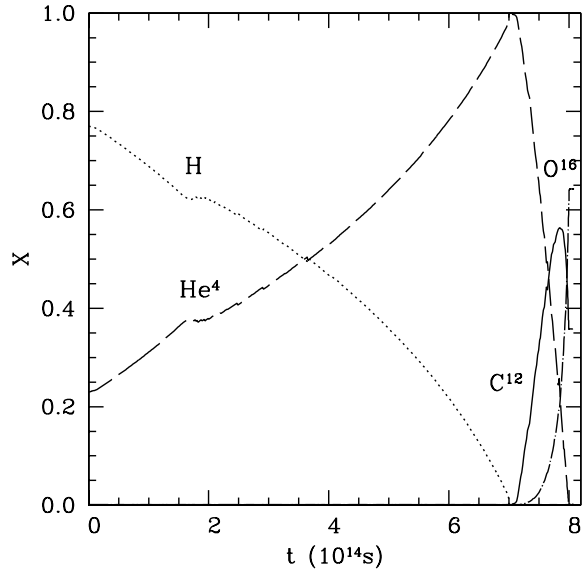}
\caption{Evolution  of  the  central  abundances of  the  9  $M_\odot$
	primary until the beginning of carbon burning.}
\end{figure}

The evolution of  the surface radius and convective  regions are shown
in the  top and bottom panels  of Fig.  3, respectively.   The part of
the  CHB phase  during which  the pp--chains  dominate can  be clearly
distinguished from  that at  which the CNO--cycle  takes over,  as the
size  of  the  convective   core  is  considerably  smaller  when  the
pp--chains  dominate.  The  pronounced  decrease of  the  size of  the
central  convective region  occuring at  $t \approx  \times 10^{14}$~s
corresponds to a time shortly after the reactions of the CNO cycle set
in.  Moreover,  during the  phase of CHB  dominated by  the pp--chains
($t\approx  1.45 \times  10^{14}$~s) the  radius of  the  star remains
constant, and slightly increases  afterwards.  Note, however, that the
surface radius of our model star does not increase significantly until
CHeB is over.

The evolution of the central abundances of the main isotopes (H, He, C
and O) as a  function of time up to carbon ignition  are shown in Fig.
4.  The  step present  in the  evolution of  the central  hydrogen and
helium  abundances between  $t\approx 1.5  \times 10^{14}$~s  and $2.0
\times 10^{14}$~s is  due to the competition between  two effects.  On
one hand,  hydrogen burning tends  to decrease its abundance.   On the
other hand, the  very fast increase in the  size of central convective
region  that  accompanies  the  activation  of the  reactions  of  the
CNO--cycle allows fresh hydrogen to be engulfed and, eventually, keeps
its abundance  constant.  The  slopes of $X(^1$H)  and $X(^4$He)  as a
function  of time are  also different  during the  phase in  which the
pp--chains dominate  and during the CNO--cycle burning  phase.  At the
end of the  helium burning phase the He--exhausted core  has a mass of
about $1.1 \, M_{\sun}$, and is composed of carbon ($X(^{12}$C$)\simeq
0.34$) and oxygen ($X(^{16}$O$)\simeq 0.65$).  These abundances differ
from  those  obtained by  Marigo  (2001),  whose calculations  yielded
$X(^{12}$C$)/X(^{16}$O$)=0.658$.  Also  the size of  the He--exhausted
core shown  here is smaller than  that obtained by  Marigo (2001), who
obtained a $1.29 \,  M_{\sun}$ He--exhausted core.  The differences in
the CO core masses and  abundance ratios are probably a consequence of
the  different  treatments  adopted  for  convection.   Marigo  (2001)
considered a  moderate amount  of overshooting in  their calculations,
whereas we are not using overshooting at all.

%_____________________________________________________________________

\section{The carbon burning phase}

\begin{figure}
\vspace{7.9cm}
\hspace{-2.7cm}
\includegraphics{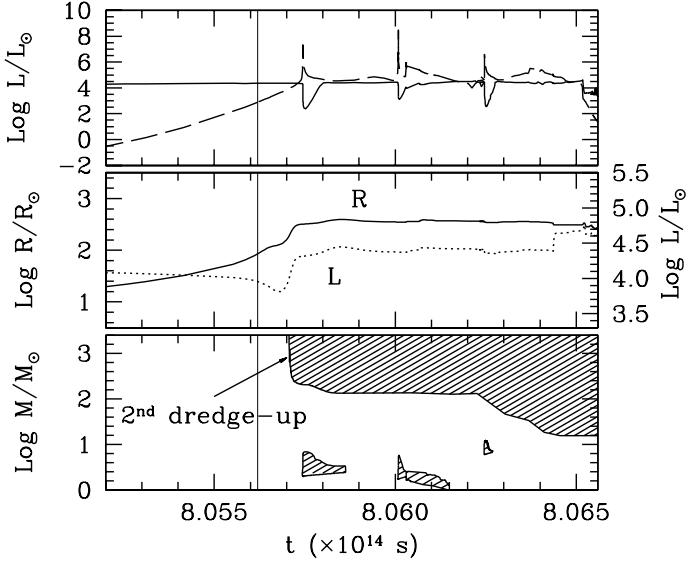}
\caption{Temporal evolution  of the main  structural parameters during
	the bulk  of the carbon burning  phase.  In the  top panel the
	luminosity  associated  to  helium  burning (solid  line)  and
	carbon  burning (dashed  line) are  shown.  The  central panel
	displays the  evolution of  the stellar surface  radius (solid
	line) and of the luminosity (dashed line). In the bottom panel
	the evolution  of the  convective regions is  represented. See
	text for details.}
\end{figure}

Figure 5  shows the temporal  evolution of main  structural quantities
just after the CHeB phase and during the bulk of the carbon phase.  In
particular,  in  the  top  panel  of  Fig.~6  the  helium  and  carbon
luminosities are  shown (solid  and dashed lines,  respectively).  The
central panel  of Fig.~5  shows the variations  of the radius  and the
luminosity.  Finally, the bottom panel of Fig.~5 shows the behavior of
the convective regions  during this phase.  The vertical  thin line in
all three panels indicates when  the carbon burning luminosity is 10\%
of the  helium burning  luminosity, corresponding to  $t=8.0562 \times
10^{14}$~s. Regions to  the left of this line  correspond to the phase
in  which the  helium burning  shell  is the  dominant nuclear  energy
source, whereas regions to the right of this line correspond to phases
in which carbon burning is significant.

Once the CHeB  phase has been completed, the  He-burning reactions set
in a  shell and, simultaneously, the  star begins an  excursion to the
red part  of the  Hertzsprung--Russell diagram at  slightly decreasing
luminosities, as shown  in Fig.~2 and in the  central panel of Fig.~5.
At this  point ($t=8.0571\times 10^{14}$~s) the surface  radius of the
star has increased from an initial  value of about $20 \, R_{\sun}$ up
to roughly  $320 \, R_{\sun}$.  This overall  expansion is accompanied
by the formation of a convective envelope, whose inner edge is able to
penetrate relatively deep inside the  star --- see the bottom panel of
Fig.~5.   However, the  bottom of  the convective  envelope  stops its
advance  at  $t=8.0571 \times  10^{14}$~s,  before  it  can reach  the
regions of the star that  have been previously enriched in products of
the H nuclear reactions.  During this process, when the maximum extent
inwards of  the base of the  convective envelope (BCE from  now on) is
reached ($M_{\rm  BCE}=2.15 \, M_{\sun}$), the  surface abundances (by
mass) of carbon, nitrogen  and oxygen are, respectively, $X(^{12}$C$)=
8.74\times   10^{-12}$,   $X(^{14}$N$)=   6.61\times   10^{-10}$   and
$X(^{16}$O$)= 7.84\times 10^{-12}$.  Therefore, during the early phase
of this  (so-called) {\sl  second} dredge--up, no  significant changes
can be  observed in the surface  composition of the star.   It must be
noted that, strictly speaking, we do not find a {\sl first} dredge--up
during hydrogen shell burning, as it is the case of higher metallicity
objects, and consequently it must be stressed that calling {\sl second
dredge--up} to this  penetration of the convective envelope  is just a
matter of keeping consistency with  the nomenclature used for stars of
larger metallicity.

Carbon burning begins just  before the second dredge--up, at $t=8.0534
\times  10^{14}$~s, while  the star  is still  climbing the  red giant
branch.  Note that  --- as previously mentioned ---  at time $t=8.0571
\times  10^{14}$~s,  the  surface  radius  stops  increasing  and  the
convective envelope temporarily halts  its inner advance.  This is due
to the increasing energy flux  released by the vigorous carbon burning
reactions  occuring   deep  inside  the  He--exhausted   core.   As  a
consequence  of the  ignition of  carbon the  He-exhausted  core stops
contracting and releasing  gravitational energy and, consequently, the
BCE stops moving  inwards. Our model star behaves in  many senses in a
way  similar to  that of  solar metallicity  stars of  slightly higher
mass.   In particular,  carbon is  burnt off-center  in  conditions of
partial degeneracy  in a series  of three consecutive flashes  --- top
panel of Fig.~5.   The first flash starts at  $M_{\rm r} \simeq 0.33\,
M_{\sun}$  --- bottom  panel  of  Fig.~5 ---  and  the carbon  burning
luminosity  reaches   values  as  high  as   $L_{\rm  C}\simeq  10^8\,
L_{\sun}$.  

\begin{figure}[t]
\vspace{9.1cm}
\hspace{-2.7cm}
\includegraphics{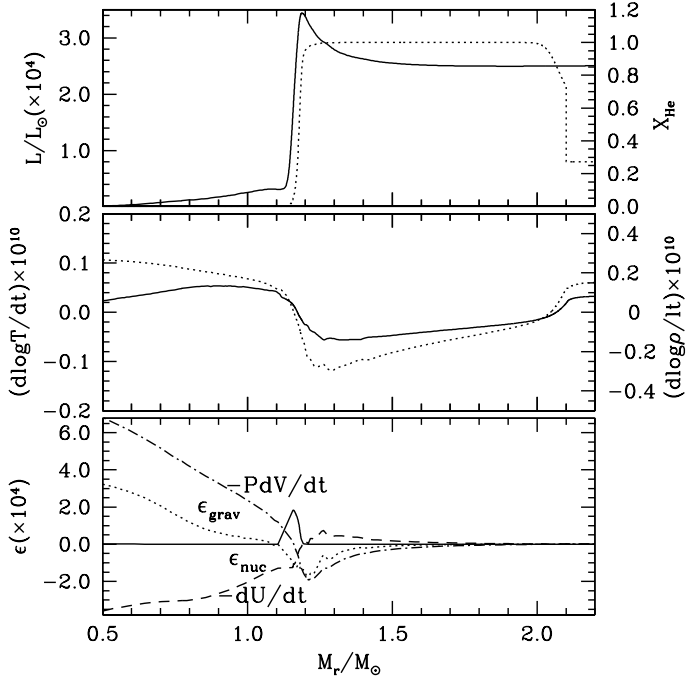}
\caption{Relevant  structural and dynamical  quantities for a model at
	time  $t=8.0616  \times  10^{14}$~s,  just  before  the second
	advance of the convective envelope, in a region extending from
	below  the  helium   burning  shell  until  the  base  of  the
	convective  envelope.  Upper panel:  velocity (solid line) and
	helium profile  (dotted  line).  Middle panel:  time variation
	of the  temperature  (solid line) and density  (dotted  line).
	Lower panel:  energy release rates (nuclear and  gravothermal)
	along with the time derivative of the internal  energy and the
	work of expansion.  The nuclear  energy  release rate has been
	divided by 10 in order to fit into the scale.}
\end{figure}

Fast injections of  energy at very high rates  and in relatively small
regions of the stellar interior induce high temperature gradients and,
consequently,  several  inner convective  regions  associated to  each
flash develop --- see the bottom  panel of Fig.~5.  Part of the energy
released in these  flashes is transformed into work  of expansion that
changes the physical  structure of the layers where  helium burning is
still active.   The expansion and cooling  of these layers  lead to an
effective decrease in helium luminosity, that recovers once the carbon
flash is finished,  very much in the same  way solar metallicity stars
of  the same  mass  do.  Nevertheless  these  dramatic changes  affect
mainly  the innermost  regions of  the star,  and they  have  a rather
limited effect on  the structure of the outer  envelope.  In fact, the
surface  radius and  luminosity remain  almost constant,  similarly to
what happens  to the $Z  = 0.02$ stars  of $10 \, M_{\sun}$,  $10.5 \,
M_{\sun}$ and $11 \,M_{\sun}$ discussed in Ritossa et al. (1995), Iben
et  al. (1997)  and Garc\'{\i}a--Berro  et al.   (1999), and  in sharp
contrast  with  the behavior  of  the  $9\,  M_{\sun}$ star  of  solar
metallicity discussed in Garc\'{\i}a--Berro et al. (1997).  The reason
of this  behavior is that  in the case  of the solar  metallicity star
carbon is ignited at  $M_{\rm r}\simeq 0.46\,M_{\sun}$, much closer to
the  helium-carbon  discontinuity  than  in  the  case  of  our  $Z=0$
model. Hence,  for the solar  metallicity star the carbon  flashes are
felt all  the way  to the  base of the  convective envelope  much more
conspicuously than in the $Z=0$ model star.

As  a  consequence  of  the  second flash,  at  time  $t=8.06021\times
10^{14}$~s,  two important  facts  occur.  First,  the carbon  burning
front reaches  the center  of the  star.  It is  worth noting  at this
point that  although the main  carbon burning phase  (corresponding to
the second flash) starts indeed as  a flash it proceeds as a flame, as
discussed in  depth in Garc\'\i  a--Berro et al. (1997).  Second, once
the center is reached, a new inward advance of the convective envelope
takes place (see  Fig.  5).  During the first  phase of the dredge--up
episode, the base of the convective envelope reached a mass coordinate
of $2.15 \,  M_{\sun}$, at which it stabilized for  most of the carbon
burning phase.  After the carbon  burning front reaches the center the
BCE moves down  to the regions located deeper in the  star.  As it can
be seen in Fig.  5, this penetration of the convective envelope occurs
in  two stages, the  first one  reaching $M_{\rm  r}=1.74\, M_{\sun}$,
where  it  stops temporarily,  and  the  second  one reaching  $M_{\rm
r}=1.20\,  M_{\sun}$.  In  fact  the  BCE gets  deep  enough to  reach
regions of the star where hydrogen and helium burning have efficiently
proceeded  and, therefore,  the  elements of  the  CNO--cycle and  the
$\alpha$-products are  dredged--up to the stellar  surface.  The final
surface  abundances   of  carbon,  oxygen  and   nitrogen  after  this
additional  dredge-up  episode  are,  respectively,  $X$($^{12}$C)$  =
2.04\times   10^{-4}$,  $X$($^{14}$N)$   =  2.41\times   10^{-6}$  and
$X$($^{16}$O)$ = 3.47\times 10^{-6}$.

\begin{figure}[t]
\vspace{9.1cm}
\hspace{-2.7cm}
\includegraphics{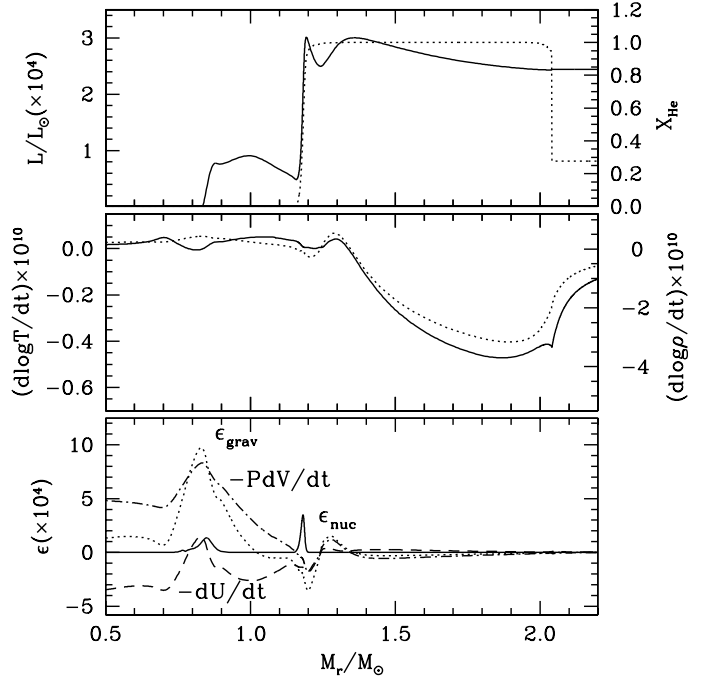}
\caption{Same as  Fig. 8, but for time  $t=8.0629 \times 10^{14}$~s,
        during the second inner advance of the envelope.}
\end{figure}

As already discussed, the first  inwards advance of the BCE took place
during the helium burning shell phase. To reach a better understanding
on how the  dredge--up process occurs, we have  represented in Figs.~6
and 7 the luminosity and  helium abundance profiles (upper panel), the
relative variation  of the temperature  and density with  time (middle
panel), and  the different rates  of energy generation  and absorption
(bottom  panel).  Fig.~6  corresponds to  a  time $t  = 8.0616  \times
10^{14}$~s, just after the carbon burning front has reached the center
of the star.  As it can be seen, the generation of nuclear energy only
occurs at the helium burning shell (HeBS), which is located at $M_{\rm
r} \simeq 1.1926\,  M_{\sun}$.  This  energy is partially  absorbed in
the region above the HeBS (of thickness $0.2 \,M_{\sun}$) and produces
a slight  expansion of these layers.   Between this zone  and the BCE,
the  structure variables  $T$  and $\rho$  and  the luminosity  remain
approximately constant.

\begin{figure}[t]
\vspace{7.9cm}
\hspace{-2.7cm}
\includegraphics{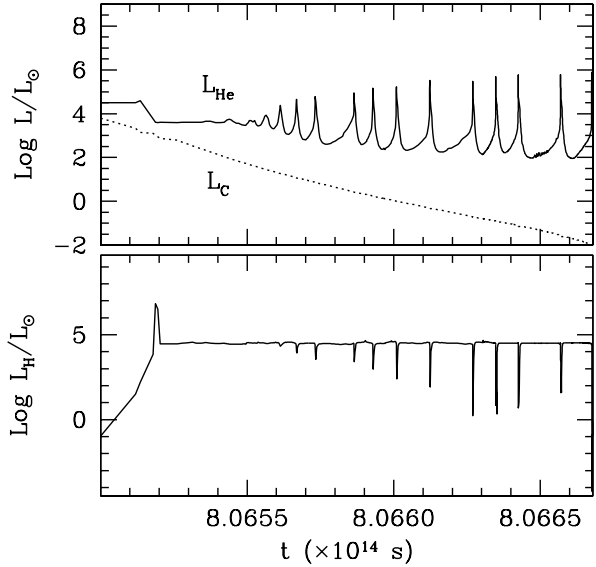}
\caption{Temporal evolution of the  first 10 thermal pulses. The upper
	panel  displays the luminosity  associated to  helium burning,
	whereas lower panel represents the evolution of the luminosity
	associated to hydrogen burning.}
\end{figure}

Fig.~7 shows  the same  quantities as in  Fig.~8, but  at a time  $t =
8.0629  \times 10^{14}$~s,  during  the second  inner  advance of  the
envelope.  The situation is now  very different since at this moment a
new   mild  carbon  flash   is  occuring   relatively  close   to  the
carbon--helium discontinuity.  The region  in which carbon is burnt is
located in the most external regions of the partially degenerate core,
at  only $0.3  \,  M_{\sun}$ below  the carbon--helium  discontinuity.
Consequently,  the  layers  between the  carbon--helium  discontinuity
(which is located at $M_{\rm  r}\simeq 1.1936\, M_{\sun}$) and the BCE
start expanding  and cooling.  Hence,  the radiative gradient  in this
region increases  over the  adiabatic gradient and,  consequently, the
BCE  advances  towards the  center.   However,  and  opposite to  what
occured in the previous carbon  flashes, the third flash is not strong
enough to  halt the  expansion and cooling  of the region  between the
HeBS and the BCE, and therefore, the convective zone keeps advancing.

As it  has been  already mentioned, after  the second carbon  flash, a
third flash of smaller strength  occurs.  The maximum luminosity is in
this case  of about $L_{\rm C}  \approx 3.0 \times  10^6 \, L_{\sun}$.
Moreover, in  this case the  associated convective zone develops  in a
region  of the  core which  is closer  to the  BCE and,  hence, mildly
degenerate.   After this flash,  the remaining  carbon burns  out more
quietly.  The subsequent  small decrease in the surface  radius of the
star (see the central panel of  Fig.~5) and the overall heating of its
envelope,  as well as  the presence  of CNO  elements in  regions just
below the base  of the convective envelope allows  a reignition of the
hydrogen  burning shell.   The conditions  of degeneracy  allow  for a
hydrogen shell  flash --- see Fig.~8.   This flash has  a magnitude of
$L_{\rm H}\simeq 10^7 \,L_{\sun}$, lasts  for 60~yr, and occurs at $t=
8.065176\times  10^{14}$~s in  a region  ($M_{\rm  r}\simeq 1.192763\,
M_{\sun}$) just  below the BCE  (which is located at  $M_{\rm r}\simeq
1.192843 \, M_{\sun}$), and very  close to the locations of the almost
extinguished  carbon burning  shell  at $M_{\rm  r}\simeq 1.178429  \,
M_{\sun}$  and  of  the  helium  burning shell,  at  $M_{\rm  r}\simeq
1.191962\, M_{\sun}$.  The associated  convective region is very small
($\sim  10^{-5}\,M_{\sun}$)  and rapidly  merges  with the  convective
envelope.   At this  time the  carbon luminosity  is  $L_{\rm C}\simeq
1.3819\times  10^3\, L_{\sun}$  and the  helium luminosity  is $L_{\rm
He}\simeq  1.1071\times  10^4\,  L_{\sun}$.  This flash  is  a  direct
consequence of  the the compression of  the layers just  below the BCE
which,  in turn,  is a  consequence of  the extinction  of  the carbon
burning  shell.   However,  after  the resurrection  of  the  hydrogen
burning shell degeneracy  is quickly removed due to  the expansion and
cooling which derive  from the fast injection of  energy in the layers
involved.  Afterwards,  hydrogen burning is not  quenched, but instead
it  proceeds at  approximately  constant luminosity  of about  $L_{\rm
H}\approx  3.0\times  10^4\,   L_{\sun}$.   The  reactivation  of  the
hydrogen  burning shell  increases  the  mass of  the  He buffer  and,
consequently, marks  the beginning of  the alternance of  hydrogen and
helium  shell flashes,  which interchange  themselves as  main nuclear
energy suppliers  of the star.   Therefore, our model star  begins the
thermally  pulsing  Super-Asymptotic   Giant  Branch,  which  will  be
described in  detail in \S 4.  The term Super-AGB stars  was coined by
Ritossa et al.  (1995) to refer  to AGB stars with ONe cores.  To this
regard it  is worth  having a  look at the  abundance profiles  of the
inner ONe  core.  The core  abundances after the  main carbon--burning
phase are  shown in  Fig.~9.  The abundance  pattern does  not present
striking   differences   with  respect   to   the  solar   metallicity
corresponding core.  However, it is  worth mentioning that the core of
the zero  metallicity star is slightly overabundant  in $^{20}$Ne with
respect to  the solar  metallicity one ---  $X(^{20}$Ne)$=0.28$ versus
$X(^{20}$Ne)$=0.24$ --- and somewhat underabundant in oxygen. Finally,
the final  mass of the C-exhausted  core in the $Z=0$  case is $M_{\rm
c}=1.15\, M_{\sun}$, $0.09 M_{\sun}$  larger than that of its $Z=0.02$
counterpart.

\begin{figure}[t]
\vspace{7.9cm}
\hspace{-2.7cm}
\includegraphics{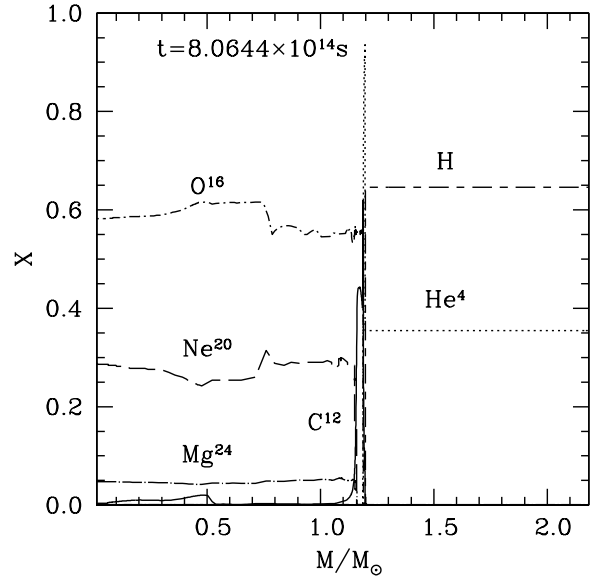}
\caption{Abundance  profiles of  the core  at  the end  of the  carbon
        burning phase.}
\end{figure}

%______________________________________________________________________

\section{The thermal pulses}

Fig.~8  shows the temporal  evolution of  the hydrogen  luminosity. We
have computed  14 thermal  pulses of which  only the first  10 thermal
pulses are  shown in Fig.~8. After  the first hydrogen  flash, a quiet
phase during which helium  accumulates follows.  During this phase the
luminosity  of  the hydrogen  burning  shell  remains nearly  constant
($L_{\rm H}\simeq 3.0 \times  10^4\, L_{\sun}$) and the helium burning
shell is  still active, opposite to  what was found by  Ritossa et al.
(1995) for  a solar metallicity  star of  $10 \,  M_{\sun}$.  However,
helium is radiatively burnt at the base of the helium buffer at a rate
smaller than  that of hydrogen at  the base of its  burning shell, the
net result being that the  mass of the helium buffer increases.  After
some  time,   the  helium  buffer  reaches  the   critical  mass  and,
consequently, a series of thermal pulses begins.  The first few pulses
are rather weak,  as usually found for this kind  of stars --- Ritossa
et  al.   (1995) and  Garc{\'\i}a--Berro  et  al.   (1997) found  some
mini-pulses, which  are totally equivalent  to what is found  here ---
but, as time advances, the amplitude of the pulses increases.  Note as
well that the inter-pulse time  interval also varies from one cycle to
the next. By  the end of the calculations shown  here the amplitude of
the  pulses remained  almost  constant with  an  approximate value  of
$L_{\rm  He}^{\rm  max}\simeq  6.0\times  10^5 \,  L_{\sun}$  for  the
maximum luminosity  and of about  $L_{\rm He}^{\rm min}\simeq  10^2 \,
L_{\sun}$  for the  minimum luminosity,  whereas the  inter-pulse time
interval  did not,  ranging between  $\tau\approx 250$~yr  and 400~yr.
For  the  sake  of  comparison,  it  is  interesting  to  recall  that
Garc\'{\i}a--Berro et al.  (1997)  found a maximum luminosity of about
$6.0  \times 10^4  \,  L_{\sun}$  for their  solar  metallicity $9  \,
M_{\sun}$  model  star,  and  a  much more  regular  inter-pulse  time
interval of  400~yr, whereas  Ritossa et al.   (1995) found  a maximum
luminosity of about  $2.7 \times 10^6 \, L_{\sun}$  and an inter-pulse
period of roughly 200~yr for  their solar metallicity $10 \, M_{\sun}$
model.   Thus the  thermal  pulses of  our  $Z=0$ model  star of  $9\,
M_{\sun}$ are similar to those of a model star of slightly larger mass
of solar metallicity, as we  have also found for other characteristics
(coupling of the  interior and the envelope during  the carbon burning
phase,  size of  the core,  chemical composition  of  the He-exhausted
core\ldots) of our model star.

The  evolution  of the  temperature  at  the  base of  the  convective
envelope during the pulses is  also midway to those of $10\, M_{\sun}$
and  $9\,  M_{\sun}$ model  stars  of  solar  composition. During  the
mini-pulses it adopts  a typical value of $T_{\rm  BCE}\simeq 3 \times
10^7$~K  monotonically  increasing  to  $T_{\rm  BCE}\simeq  8  \times
10^7$~K  up to  the  eigth pulse.   From  this thermal  pulse on,  the
temperature  at the base  of the  convective envelope  remains roughly
constant and enough  to transform part of the  freshly synthesized CNO
elements into $^{25}$Mg and neutrons.   As to the existence of a third
dredge--up, we  have not  detected such a  process after  following 14
helium flashes.  However, it is  important to realize at this point of
the discussion that dredge--up processes are strongly dependent on the
treatment of convection, and the  fact that we have not introduced any
overshooting in our calculations  might have prevented the development
of an additional dredge--up.
                                  
It  is  worth recalling  at  this point  of  the  discussion that  the
thermally pulsing  AGB phase of  low and intermediate  mass primordial
stars has been thoroughly studied by Siess et al.  (2002).  They found
two different mechanisms allowing for the thermal pulses.  For objects
between 1  and $5 \,  M_{\sun}$, a convective instability  develops at
the  He--H   discontinuity  causing  the  ingestion   of  carbon  and,
consequently,  reactivating the  CNO--cycle.  Instead,  for the  $7 \,
M_{\sun}$ object  it is a  previous dredge--up phase which  allows the
enrichment in  CNO products and  the subsequent behaviour as  a normal
thermally pulsing  AGB star.  Our  $9 \, M_{\sun}$ model  star behaves
very much  in the same  way as the  $7 \, M_{\sun}$ star  described by
Siess et  al.  (2002).   Both our  results and those  of Siess  et al.
(2002) contradict  the pioneering  work  of Fujimoto  et al.   (1984).
These authors found that there was a threshold value at $M_* = 0.73 \,
M_{\sun}$  such  that stars  whose  core  mass  was lower  than  $M_*$
experienced thermal  pulses, but stars  with a core more  massive than
this  threshold value could  only experience  thermal pulses  if their
metallicity was above $Z  \approx 10^{-7}$.  However, as already shown
by  Siess et  al.  (2002)  and  here, the  pollution in  CNO--elements
caused  by convection  in the  He--H discontinuity  zone or  caused by
previous   dredge--up  phases  makes   this  requirement   on  initial
metallicity unnecessary.

Finally, it  is interesting to  have a look  at the surface  ratios of
carbon, nitrogen and oxygen. For the solar metallicity stars of masses
9 and  $10\, M_{\sun}$  studied before in  Ritossa et al.   (1995) and
Garc\'\i  a--Berro  et  al.   (1997)  these  ratios  are  respectively
(C:N:O)=(1.63:3.63:6.93)   and  (2.35:4.25:6.26)   after   the  second
dredge-up,  and the  total metallicities  are in  both  cases $Z\simeq
0.012$ at this  moment. In sharp contrast the  initially $Z=0$ star of
$9\,M_{\sun}$  studied here  presents  (C:N:O)=(2.03:0.02:0.03) for  a
total metallicity of  $Z=2.1\times 10^{-4}$. Thus, observationally our
model star would look like a C-enhanced metal-poor star.

%_____________________________________________________________________

\section{Mass loss and evolution to white dwarf}

In   Fig.~10   the   latest   stages   of   the   evolution   in   the
Hertzprung-Russell  diagram of  the zero  metallicity $9  \, M_{\sun}$
star are presented.  The  most important evolutionary stages have been
highlighted  along   the  evolutionary  track.    In  particular,  the
beginning of the  carbon burning phase has been  marked with a circle,
whereas the diamond  shows the end of the  thermally pulsing Super-AGB
(TP-SAGB) phase.   After the eleventh thermal pulse,  when the surface
radius  is  $R_{\rm  s}  \simeq  320  \,  R_{\sun}$  and  the  surface
luminosity is $L_{\rm  s} \simeq 6.5\times 10^3 \,  L_{\sun}$, we have
assumed, somehow arbitrarily, that a mass-loss episode ensues.  Before
the eleventh pulse  no mass-loss was assumed to  occur. This mass-loss
episode is clearly marked in Fig.~10 using a thick solid line.

As it occurs for solar metallicity objects of analogous ZAMS mass, the
mass-loss episode  is plagued with many uncertainties,  as no definite
mass-loss rates are available for  low metallicity stars, and not even
there  is a  well studied  mechanism responsible  for  these mass-loss
episodes --- see, however, Marshall  et al. (2004).  Stellar winds for
the case of  single stars as well as Roche lobe  overflow for the case
of binaries are  likely to play important roles,  as it usually occurs
for solar metallicity  stars of analogous masses.  With  the values of
the luminosity and the radius at the end of the eleventh thermal pulse
the mass-loss  rates due to stellar  winds would be  significant for a
solar    metallicity    star,    of    the   order    of    $10^{-6}\,
M_{\sun}$~yr$^{-1}$, according to Reimers (1975) and Nieuwenhuijzen \&
de Jager (1990).  We recall  that the convective envelope of our model
star  presents  a   significant  enrichment  in  metals  $(Z=2.1\times
10^{-4})$  which,  nevertheless,  is  smaller  than that  of  a  solar
metallicity  star  of  the   same  mass.   Consequently,  to  a  first
approximation we can consider the mass-loss rates of Nieuwenhuijzen \&
de  Jager  (1990)  as  representatives  of the  real  mass-loss  rate.
Additionally,  the extension  of the  mass-loss episode  must  also be
taken  with some  caution and,  consequently, the  final limit  of the
thick line in Fig.~10 must also be considered as an approximate value.

\begin{figure}
\vspace{7.3cm}
\hspace{-2.7cm}
\includegraphics{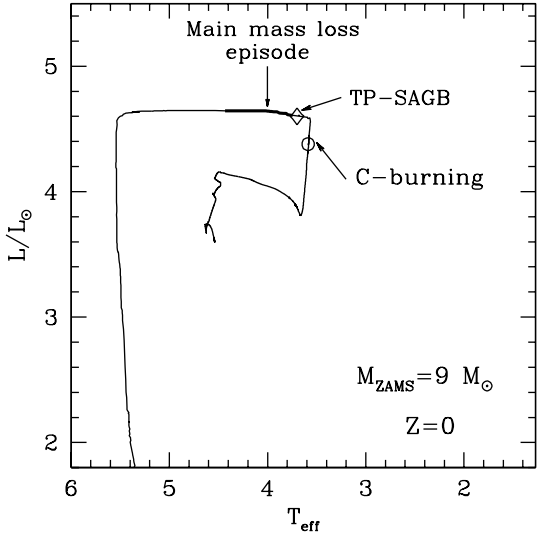}
\caption{Evolution  in the  Hertzsprung-Russell diagram  of the  $9 \,
	M_{\sun}$  star of zero  metallicity (solid  line) and  of the
	$Z=0.02$  star of  the same  mass (dotted  line).   The circle
	corresponds to  the beginning of  the C-burning phase  and the
	diamond  corresponds  to  the  end of  the  thermally  pulsing
	Super-AGB  (TP-SAGB)  phase.   The  thick line  at  the  early
	horizontal  track corresponds  to the  mass loss  episode. See
	text for further details.}
\end{figure}

Keeping this in mind, we have performed different sets of calculations
of the latest  evolutionary stages, to cover the  effects of mass loss
with  rates between  $\sim  10^{-7} \,  M_{\sun}$~yr$^{-1}$ and  $\sim
10^{-4}\, M_{\sun}$~yr$^{-1}$, which  bracket the representative value
derived  above.  The  evolutionary sequences  computed with  low rates
would describe the effects of mass  loss due to stellar winds, and the
sets  of evolutionary  sequences computed  with higher  rates (between
$10^{-5}     \,     M_{\sun}$~yr$^{-1}$     and    $\sim     10^{-4}\,
M_{\sun}$~yr$^{-1}$)  might  describe  mass  loss due  to  Roche  lobe
overflow  in  a  binary  system.   The mass-loss  rate  used  for  the
evolutionary sequence  shown in Fig.~10  is $\dot M=3.0\times10^{-6}\,
M_{\sun}$~yr$^{-1}$.  Finally, it is important to say that, in all the
cases studied here we have encountered some computational difficulties
at the latest  evolutionary stages --- as it is  usually found for AGB
stars  with  carbon-oxygen  cores   with  masses  smaller  than  those
considered here --- but no significant differences regarding the final
size and composition of the remnant core have been found.

\begin{figure}
\vspace{7.3cm}
\hspace{-2.7cm}
\includegraphics{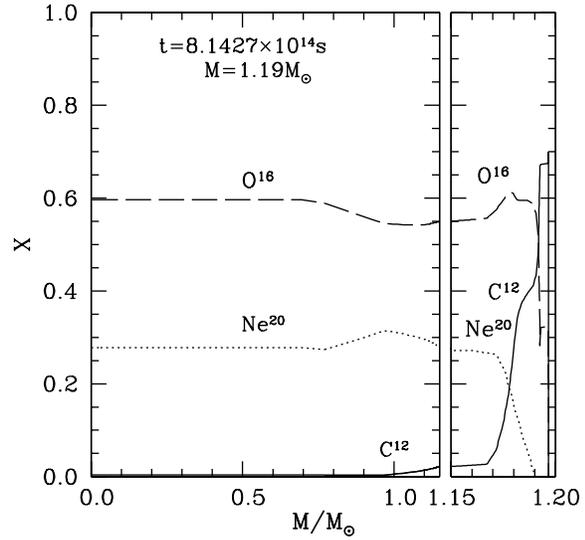}
\caption{Main isotope  abundance profile of the  zero metallicity $9\,
        M_{\sun}$ remnant star.}
\end{figure}

Finally, Fig.~11  shows the  chemical profiles of  the remnant  of the
zero metallicity $9  \, M_{\sun}$ star. The mass of  the core is $\sim
1.19\, M_{\sun}$.  If we compare with the central regions of the model
presented in Fig.~9, corresponding to  a time shortly after the end of
the carbon burning  phase, it can be seen that  some sharp features in
the abundance profiles after the bulk of the carbon burning phase have
been smeared out due  to the Rayleigh-Taylor instabilities (Salaris et
al., 1998)  that occur  in the subsequent  evolution of the  star, but
neither  the  size  nor  the  general composition  of  the  core  have
substantially  changed. The  main  isotopes present  in  the core  are
$^{16}$O  and $^{20}$Ne, with  mass abundances  $\sim 0.59$  and $\sim
0.28$, respectively. As in the  case of the solar metallicity stars of
similar mass  (Ritossa et al.,  1995; Gil--Pons \&  Garc\'\i a--Berro,
2001), some small amount of  unburnt carbon is present (with a maximum
mass abundance of $\sim 0.02$ before the Rayleigh-Taylor instabilities
mix it  through out the inner  regions of the core).   Finally, in the
$Z=0$ case no significant amounts of $^{24}$Mg nor $^{23}$Na have been
found.

%_____________________________________________________________________

\section{Summary and Conclusions}

We  have computed the  evolution of  a zero  metallicity $9\,M_{\sun}$
star from  the zero age  main sequence until  the formation of  an ONe
white  dwarf,  through core  carbon  burning  and  the TP-SAGB  phase.
Previous works  on zero  metallicity intermediate mass  evolution have
recently been made  by Marigo et al.  (2001),  Marigo (2002), Heger \&
Woosley (2002) and  Chieffi et al.  (2001), but  these authors did not
follow carbon  ignition in  the degenerate core.   Hence, this  is the
first calculation  in which the  carbon burning phase and  the thermal
pulses of  a heavy--weight intermediate mass primordial  star has been
followed self-consistently.

Our results compare  favorably with those of Marigo  et al. (2001) and
Marigo (2002) for a $9\,M_{\sun}$  model star. In particular, the time
scales for hydrogen burning  are notably similar in both calculations.
We do not  obtain such a nice  agreement for the final size  of the CO
cores, since our calculations  yield a $\simeq 1.1\,M_{\sun}$ CO core,
whereas they obtain a CO core of $1.28\,M_{\sun}$.  However, it should
be  noted  that even  though  we both  are  using  the OPAL  radiative
opacities (Rogers  \& Iglesias, 1992;  Iglesias \& Rogers,  1993), and
the conductive  opacities of Hubbard  \& Lampe (1969),  our respective
equations of state  are not exactly the same.   Concerning the nuclear
reaction  rates, we  both are  using basically  the reaction  rates of
Caughlan \& Fowler (1988), but Marigo et al. (2001) increased the rate
for  the controversial  $^{12}$C$(\alpha,\gamma)^{16}$O reaction  by a
factor 1.7  (Weaver \& Woosley,  1993), whereas we did  not.  Finally,
the factor that is more likely to ultimately determine the differences
in the sizes of our respective  CO cores, is probably the treatment of
convection.   Marigo  et al.   (2001)  used  moderate convective  core
overshooting, whereas we did not use it at all.

We obtain  a partially degenerate CO  core in which  carbon is ignited
off-center. Carbon burning proceeds  through a series of flashes which
generates  inner convective  regions,  as  it is  found  for stars  of
similar mass of solar  metallicity.  Regarding convection, it is worth
noticing at  this point  that a calibration  of the parameters  of the
Mixing Length Theory (and in particular of the amount of overshooting)
can only be  made for the case of solar  metallicity by comparing with
the   abundant  observational   material.   This   procedure  provides
reliability to the  MLT results, but the problem in  the case of $Z=0$
stars is  the lack of  observations and, therefore, of  an appropriate
calibration.   Given  the  lack  of  observational  material  we  have
preferred to stick  as close as possible to  our previous calculations
(Gil--Pons \& Garc\'\i a--Berro,  2001; 2002) in which no overshooting
was  adopted.  Although the  evolution during  carbon burning  for the
$Z=0$ model star  is very similar to that of  solar metallicity, it is
important to note one difference between the two models.  Namely, that
the envelope  of the $Z=0$ star  behaves as if it  were more decoupled
from  the core  than its  solar metallicity  counterpart.   Hence, its
structure is not  significantly changed by the release  of energy that
occurs at the innermost regions  of the star.  Once carbon is depleted
in most of the central regions  of the He--exhausted core we find that
the base  of the convective  envelope advances inwards and  dredges up
nuclear processed  elements.  Finally,  the hydrogen burning  shell is
reactivated and a thermally pulsing Super-AGB (an AGB star with an ONe
core) is  formed.  We have  followed the first  11 pulses and  we have
found that their characteristics  are midway between to those computed
for solar metallicity stars of 9 and $10\,M_{\sun}$.

Somehow arbitrarilly we have chosen  mass loss to begin at $t=8.066307
\times 10^{14}$~s, when the eleventh thermal pulse is already over and
the surface radius is approximately $320 \, R_{\sun}$.  At this stage,
the dredge--up  process following  core helium exhaustion  has already
taken place.   We have not delved  into the causes of  mass loss, that
might correspond either to stellar winds or to Roche lobe overflow (in
binary  systems).   Instead,   because  no  parametrization  of  these
processes  has been  made up  to  now, we  have computed  a series  of
sequences with mass  loss rates between $10^{-7} \,M_{\sun}$~yr$^{-1}$
and $\sim 10^{-4} \,M_{\sun}$~yr$^{-1}$.   In any case, no significant
differences have been found regarding  the mass of the remnant and its
composition.  We  obtain a ultramassive  helium-rich $1.19\, M_{\sun}$
white dwarf  formed by  a $1.17 \,  M_{\sun}$ oxygen--neon core  and a
$0.02  \, M_{\sun}$  carbon--oxygen buffer.   Such  ultramassive white
dwarfs have been detected by  Finley et al.  (1997) in a spectroscopic
survey of DA white dwarfs hotter than $\sim 25000$~K, and by Dupuis et
al.  (2002) in the EUVE  and ROSAT surveys.  More recently, Liebert et
al.  (2004)  have found  a tail  of massive white  dwarfs in  the mass
distribution of  white dwarfs in  the Palomar Green  Survey. Moreover,
these authors  have pointed out that  of 28 white dwarfs  all but nine
are within  $\pm30^\circ$ of the Galactic plane,  thus suggesting that
there is an association between these white dwarfs and that of B and O
stars.  Additionally, they noticed  that the ultramassive component of
the white  dwarf mass distribution has small  cooling ages, suggesting
massive  progenitors. In summary,  all these  surveys have  provided a
handful of ultramassive white dwarfs  and it is interesting to realize
that the derived  mass distributions show a high  mass shoulder with a
peak  centered at  $\sim 0.9\,  M_{\sun}$ that  provides an  excess of
white dwarfs  with masses between  $1.0$ and $1.2\,  M_{\sun}$.  Since
the  theoretical mass  distributions using  a  steady initial-to-final
mass relation do  not predict such behavior, Weidemann  (2000) was the
first to  ascribe this behavior  to an unsteady  initial-to-final mass
relation which  separates white  dwarfs with carbon-oxygen  cores from
those  with oxygen-neon  cores like  the one  obtained in  the present
work.  Although the unsteady behavior has also been ascribed to binary
evolution, the  matter still  remains as an  open question,  as recent
Smoothed Particle Hydrodynamics simulations (Guerrero et al., 2004) do
not totally discard the formation of an ultramassive white dwarf.  Our
simulations, thus,  reinforce the idea that  ultramassive white dwarfs
could  be  the  end-product   of  the  standard  evolution  of  single
heavy-weight intermediate mass stars.

\begin{acknowledgements}
Part of this work was supported by the MCYT grant AYA04094--C03-01, by
the European Union FEDER funds, and  by the CIRIT.  We would also like
to  acknowledge  the advise  of  our  anonymous  referee for  valuable
comments  and criticisms.  One  of us  (E.G.)  acknowledges the  Japan
Society  for  Promotion of  Science  for  a  fellowship to  visit  the
Hokkaido University, where this project was started.
\end{acknowledgements}

%_____________________________________________________________________

%_______________________________________________________________________


\begin{thebibliography}{1}

\bibitem{ADSL}  Abia, C., Dom\'{\i}nguez,  I., Straniero, O., Limongi,
        M., Chieffi, A., \& Isern, J., 2001, ApJ, {\bf 557}, 126
\bibitem{AH}  Alpher,  R.A., \& Herman,  R., 1950,  Rev.  Mod.  Phys.,
        {\bf 22}, 153
\bibitem{BN:84} Bessel, M.S., \& Norris, J., 1984, ApJ, {\bf 285}, 622
\bibitem{C:81} Carlberg, R.G., 1981, MNRAS {\bf 197}, 1021
\bibitem{C} Canuto, V., 1970, ApJ, {\bf 159}, 641
\bibitem{CAB} Carr, B.J., Bond, J.R., \& Arnett, W.D., 1983, ApJ, {\bf
        277}, 445
\bibitem{CC93}  Cassisi, S., \& Castellani,  V., 1993, ApJSS {\bf 88},
        509
\bibitem{CF}  Caughlan,  G.R., \& Fowler,  W.A., 1988,  Atom.  Data \&
        Nuc.  Data Tables, {\bf 40}, 283
\bibitem{CDLS}   Chieffi,  A., Dom\'{\i}nguez,  I.,  Limongi,  M.,  \&
        Straniero, O., 2001, ApJ, {\bf 554}, 1159
\bibitem{CS} Cox, A.N., \& Stewart, J.N., 1970, ApJSS, {\bf 19}, 243
\bibitem{CBB} Christlieb, N.,  Bessell, M.S., Beers, T.C., Gustafsson,
        B.,  Korn, A., Barklem,  P.S., Karlsson,  T., Mizuno--Wiedner,
        M., \& Rossi, S., 2002, Nature, {\bf 419}, 904
\bibitem{DAM:82} D'Antona 1982, A\&A, {\bf 115}, L1
\bibitem{DEA02} Dupuis, J., Vennes, S., \& Chayer, P., 2002, ApJ, {\bf
        580}, 1091
\bibitem{EZ}  Ezer, D., \& Cameron, A.G.W., 1971, Astrophys.  \& Space
        Sci., {\bf 14}, 399
\bibitem{FKB} Finley, D.S., Koester, D., \& Basri, G., 1997, ApJ, {\bf
        488}, 375
\bibitem{FCZ75}  Fowler, W.A.,  Caughlan,  G.R., \& Zimmermann,  B.A.,
        1975, ARA\&A, {\bf 13}, 69
\bibitem{FICT}   Fujimoto,   M.Y.,  Iben,  I.Jr.,   Chieffi,   A.,  \&
        Tornamb\'{e}, A., 1984, ApJ, {\bf 287}, 749
\bibitem{P3}  Garc\'{\i}a--Berro,  E., Ritossa, C., \& Iben, I., 1997,
        ApJ, {\bf485}, 765
\bibitem{GG:2001}  Gil--Pons,  P., \&  Garc\'{\i}a--Berro,  E.,  2001,
        A\&A, {\bf375}, 87
\bibitem{GG:2002}  Gil--Pons,  P., \&  Garc\'{\i}a--Berro,  E.,  2002,
        A\&A, {\bf396}, 789
\bibitem{GS} Gorieli, S., \& Siess, L., 2001, A\&A, {\bf 378}, 25
\bibitem{GEA04} Guerrero, J., Garc\'\i a--Berro, E., \& Isern, J., 
        2004, A\&A, {\bf 413}, 257
\bibitem{HWa} Heger, A., \& Woosley, S.E., 2002, ApJ {\bf 567}, 532
\bibitem{HWW} Heger, A.,  Woosley, S.E., \& Waters, R.,  2000, in {\sl
        ``The  First Stars''},  Eds.   A.  Weiss,  T.G.   Abel, \&  V.
        Hill, (Springer: Berlin), 121
\bibitem{HWb} Heger, A., \& Woosley, S.E., 2002, in Proc.  of $11^{\rm
        th}$ Workshop on Nucl.  Astrophys., Ringberg Castle., Germany,
        Eds.  E.  M\"{u}ller \& W.  Hillebrandt, MPA Procs., Garching,
        8
\bibitem{HL} Hubbard, W.B., \& Lampe, M., 1969, ApJSS, {\bf 18}, 297
\bibitem{P4} Iben, I., Ritossa, C., \&  Garc\'{\i}a--Berro,  E., 1997,
        ApJ, {\bf489}, 772
\bibitem{IR:93} Iglesias, C.A., \& Rogers, F.J., 1993, ApJ, {\bf 412},
        752
\bibitem{LBH}  Liebert,  J., Bergeron,  P.,  \&  Holberg, J.B.,  2004,
        ApJSS, in press, {\tt astro-ph/0406657}
\bibitem{LSC} Limongi, M., Straniero, O., \& Chieffi, A., 2000, ApJSS,
        {\bf 129}, 625
\bibitem{MGCW} Marigo, P., Girardi, C., Chiosi, C., \& Wood, R., 2001,
        A\&A, {\bf 371}, 152
\bibitem{M02} Marigo, P., 2002, A\&A, {\bf 387}, 507
\bibitem{MCK} Marigo, P., Chiosi, C., \& Kudritzki,  R.P., 2003, A\&A,
        {\bf 399}, 617
\bibitem{MEA04} Marshall,  J.R., van  Loon, J.T., Matsuura,  M., Wood,
        P.R., Ziljstra, A.A., \& Whitelock, P.A., 2004, \mn, in press
\bibitem{NU:2001}  Nakamura,  F., \& Umemura, M., 2001, ApJ {\bf 548},
        19
\bibitem{NJ}  Nieuwenhuijzen,  H.,  \&   de  Jager,  C.,  1990,  A\&A,
        {\bf231}, 134
\bibitem{PSS}  Palla, F., Salpeter, E.E., \& Stahler, S.W., 1983, ApJ,
        {\bf 271}, 632
\bibitem{R:75} Reimers, D.,  1975, Mem. Soc. R. Sci.  Li\`ege, ser. 6,
        {\bf 8}, 369
\bibitem{P2} Ritossa, C.,  Garc\'{\i}a--Berro,  E., \& Iben, I., 1995,
        ApJ, {\bf460}, 489
\bibitem{P5} Ritossa, C.,  Garc\'{\i}a--Berro,  E., \& Iben, I., 1999,
        ApJ, {\bf515}, 381
\bibitem{RI:92} Rogers, F.J., \& Iglesias, C.A., 1992, {\bf 401}, 361
\bibitem{SN} Saio, H., \& Nomoto, K., 1998, ApJ, {\bf500}, 388
\bibitem{S98} Salaris,  M., Dom\'\i nguez, I.,  Garc\'\i a--Berro, E.,
        Hernanz, M.,  Isern, J., \& Mochkovitch, R.,  1998, \apj, {\bf
        486}, 413
\bibitem{SLL}  Siess, L., Livio, M., \& Lattanzio,  J., 2002, ApJ {\bf
        570}, 329
\bibitem{TS00} Tumlinson, J., \& Shull, M., 2000, ApJL {\bf 528}, 65
\bibitem{WW} Weaver, T.A., Woosley, S.E., 1993, Ph.  R., {\bf227}, 65
\bibitem{W00} Weidemann, V., 2000, A\&A, {\bf 363}, 647
\bibitem{YS:86} Yoshii, Y.  \& Saio, H., 1986, ApJ {\bf 301}, 587

\end{thebibliography}
\end{document}